\documentstyle[twocolumn,prc,aps,psfig]{revtex}
\begin{document}\hbadness=10000
\twocolumn[\hsize\textwidth\columnwidth\hsize\csname %
@twocolumnfalse\endcsname
\title{On hadron production in Pb--Pb collisions at 158$A$ GeV}
\author{Johann Rafelski  and Jean Letessier}
\address{
Department of Physics, University of Arizona, Tucson, AZ 85721\\
and\\
Laboratoire de Physique Th\'eorique et Hautes Energies\\
Universit\'e Paris 7, 2 place Jussieu, F--75251 Cedex 05.
}
\date{February 24, 1999; September 6, 1999}
\maketitle
\begin{abstract}
A Fermi statistical model analysis of hadron abundances and spectra
obtained in several relativistic heavy ion collision experiments 
is utilized to characterize a particle source. 
 Properties consistent with  a disintegrating,  hadron evaporating,
deconfined quark-gluon plasma phase fireball are obtained, with
a baryochemical potential $\mu_{B}=200$--210\,MeV, and a temperature 
$T_f\simeq 140$--150\,MeV, significantly 
below previous expectations.
\end{abstract}
\pacs{PACS: 25.75.-q, 25.75.Dw, 25.75.Ld, 12.38.Mh}
\vspace{-0.1in}
]
\begin{narrowtext}
Discovery and study of quark-gluon plasma (QGP), a
state  consisting of mobile, color charged quarks and gluons, is the
objective of the relativistic heavy ion research
program underway at Brookhaven National Laboratory, New York and
at CERN, Geneva \cite{QGP}. Thermalization of the constituents of
the deconfined phase created in high energy large nuclei
collisions is a well working  hypothesis, as we shall see.
The multi-particle  production processes in
158$A$ GeV Pb--Pb collisions carried out at CERN-SPS
will be analyzed in this paper, using the principles of the statistical
Fermi model \cite{Fer50}: strongly interacting particles are produced
with a probability commensurate with the size of accessible
phase space. Since the last comprehensive review  of such analysis 
has appeared \cite{Sol97}, the Pb-beam experimental 
results became available, and model improvements have 
occurred: we implement  refinements
in the phase space weights that allow a
full characterization of the chemical non-equilibria with
respect to strange and light quark flavor 
abundances \cite{Raf91,LRa99}. Consideration of the light quark 
chemical non-equilibrium is necessary in order to arrive at
a  consistent interpretation of the experimental 
results of both the wide acceptance NA49-experiment
\cite{Ody98,Puh98,Bor97,App98,Mar99} and central rapidity 
(multi)strange (anti)baryon  
WA97-experiment \cite{Hol97,Kra98,WA97}.

We further consider  here the influence  
of collective matter flow on $m_\bot$-particle spectra and
particle multiplicities obtained in a limited phase space 
domain. The different flow schemes  
have been described before \cite{Hei92}.  We adopt a radial
expansion model and  consider the causally disconnected domains
of the dense matter fireball  to be synchronized by the
instant of collision. We subsume that the particle (chemical) 
freeze-out occurs at the surface of
the fireball, simultaneously in the CM frame, but not necessarily
within a short instant of CM-time. Properties of the dense 
fireball as determined in this approach offer clear
 evidence that a QGP disintegrates  at $T_f\simeq$\,144\,MeV,
corresponding to energy density 
$\varepsilon=\cal O$(0.5) GeV/fm$^3$ \cite{Kar98}.
Our initial chemical non-equilibrium results without flow  have
been suggestive that this is the case\cite{LRPb98}, showing 
a reduction of the 
chemical freeze-out temperature from $T_f=180$\,MeV \cite{Bec98};
an earlier analysis  could not exclude yet higher hadron formation 
temperature of 270\,MeV \cite{Let97}. 

The here developed model offers a natural understanding of 
the systematic behavior of the $m_\bot$-slopes which differs
from other interpretations. 
The near equality of (inverse) slopes of nearly all strange
baryons and antibaryons  arises here by means of the  
sudden hadronization at the surface of an
exploding QGP fireball. In the hadron based microscopic 
simulations this behavior of  $m_\bot$-slopes can also
arise allowing for particle-dependent freeze-out times \cite{HSX98}.

In the analysis of hadron spectra we employ 
methods developed in  analysis  of the 
lighter  200$A$ GeV S--Au/W/Pb system  \cite{LRa99}, where  the description 
of the phase space accessible to a hadronic  particle
in terms of the  parameters we employ is given. 
Even though we use six parameters to characterize the hadron 
phase space at chemical freeze-out, there are only two truly 
unknown properties: the chemical freeze-out temperature $T_{f}$ and  
light quark fugacity $\lambda_q$\, 
(or equivalently, the baryochemical potential 
$\mu_B=3\,T_{f}\ln \lambda_q$) -- we recall that the parameters 
$\gamma_i,\,i=q,s$ controls overall
abundance of  quark pairs, while $\lambda_i$ controls the difference
between quarks and anti-quarks of given flavor. The  four other
parameters are not arbitrary, and we could have used their
tacit and/or computed values:\\
 1) the strange quark fugacity $\lambda_s$ is usually
fixed by the  requirement that strangeness balances 
$\langle s-{\bar s}\rangle=0$\, \cite{Raf91}. The Coulomb distortion
of the  strange quark phase space plays an important role in the
understanding of this constraint for Pb--Pb collisions,
see Eq.\,(\ref{lamQ}) \cite{LRPb98};\\
 2) strange quark phase space occupancy 
$\gamma_s$ can be computed  within the established kinetic theory
framework for strangeness production \cite{RM82,acta96};\\
 3) the tacitly assumed equilibrium phase space occupancy of light quarks 
$\gamma_q=1$\,; and \\
 4) assumed collective expansion to proceed at
the relativistic sound velocity, $v_c=1/\sqrt{3}$ \cite{acta96}.\\
However, the rich particle data basis allows us to find from experiment
the actual values of these four parameters, allowing to confront
the theoretical results and/or hypothesis with experiment. 

The value of $\lambda_s$ we obtain from the strangeness conservation 
condition $\langle s-{\bar s}\rangle=0$\ in QGP is, to a very good 
approximation \cite{LRPb98}:
\begin{equation}\label{tilams}\label{lamQ}
\tilde\lambda_s\equiv \lambda_s \lambda_{\rm Q}^{1/3}=1\,,\qquad
\lambda_{\rm Q}\equiv
\frac{\int_{R_{\rm f}} d^3r e^{\frac V{T}} } {\int_{R_{\rm f}} d^3r}\,.
\end{equation}
 $\lambda_{\rm Q}<1$  expresses the Coulomb deformation of 
strange quark phase space. This effect is
relevant in central Pb--Pb interactions, but not in 
S--Au/W/Pb reactions. $\lambda_{\rm Q}$ is not a fugacity that 
can be adjusted to satisfy a chemical condition,
since consideration of $\lambda_i,\ i=u,d,s$ exhausts all available
chemical balance conditions for the abundances of hadronic particles. 
The subscript ${R_{f}}$ in Eq.\,(\ref{lamQ}) reminds us 
that the classically
allowed region within the dense matter fireball is included in
the integration over the  level density.
 Choosing $R_{\rm f}=8$\,fm, $T=140$\,MeV,
$m_s=200$\,MeV (value of $\gamma_s$ is practically irrelevant),
for $Z_{\rm f}=150$ the value is $\lambda_s=1.10$\,.

In order to interpret particle abundances measured
in a restricted phase space domain, we study abundance ratios 
involving what we call compatible hadrons: 
these are particles likely to be impacted in 
a similar fashion by the not well understood 
collective flow dynamics in the fireball. 
The available particle yields are listed 
in table~\ref{resultpb2}, top section from the
experiment WA97 for $p_\bot>0.7$ GeV within a narrow
$\Delta y=0.5$ central rapidity window. Further below 
are shown  results from the large  acceptance experiment NA49, 
extrapolated to full $4\pi$ phase space coverage. 
There are 15 experimental results. The total error 
$\chi^2_{\rm T}\equiv\sum_j({R_{\rm th}^j-R_{\rm exp}^j})^2/
({{\Delta R _{\rm exp}^j}})^2$ for the  four theoretical 
columns is shown at the bottom of table~\ref{resultpb2}
along with the number of data points `$N$', parameters `$p$' 
used and  (algebraic) redundancies `$r$' connecting the 
experimental results. For $r\ne 0$ it is more appropriate 
to quote the total  $\chi^2_{\rm T}$, with a initial qualitative
statistical relevance condition  $\chi^2_{\rm T}/(N-p)<1$.
The four theoretical columns refer to results with
collective velocity $v_c$ (subscript $v$) 
or without ($v_c=0$). We consider data including 
`All' data points, and also analyze data 
excluding from analysis four $\Omega,\,\overline\Omega$ 
particle ratios, see columns marked `No-$\Omega$. 
Only in letter case we obtain a highly 
relevant data description. Thus to describe
the  $\Omega,\,\overline{\Omega}$ yields we need
an additional particle production mechanism 
beyond the statistical Fermi model. 
We noted the special role of these particles, despite
bad statistics, already in the analysis of the 
S-induced reactions \cite{LRa99}.

Considering results obtained with and without flow
reveals that the presence of the parameter $v_c$ already
when dealing only with particle abundances improves 
our ability to describe the 
data. However, $m_\bot$ spectra offer another independent measure 
of the collective flow $v_c$: for a given pair of values
 $T_{f}$ and $v_{\rm c}$ we evaluate the resulting
$m_\bot$ particle spectrum and analyze it using the spectral shape
and kinematic cuts employed by the experimental groups.
To find the best values we consider just one `mean' strange baryon
experimental  value ${\bar T}_{\bot}^{\rm Pb}=260\pm10$,
since within the error the high $m_\bot$ strange (anti)baryon
inverse slopes are overlapping. Thus when considering $v_c$ along with
${\bar T}_{\bot}$ we have one parameter and one data point more. 
Once we find best values of $T_{\rm f}$
and $v_{\rm c}$, we  study again the inverse slopes of 
individual  particle spectra. We obtain an acceptable
agreement with the experimental $T_{\bot}^j$ as 
shown in left section of table~\ref{Tetrange2}\,.
For comparison, we have also considered in the same framework the
S-induced reactions, and the right section of 
table~\ref{Tetrange2} show a  good
agreement with the WA85 experimental data \cite{WA85slopes}.
We used here as the `mean' experimental slope data point
${\bar T}_{\bot}^{\rm S}=235\pm10$. We can see that
within a significantly smaller error bar, we obtained an accurate
description of the $m_\bot^{\rm S}$-slope data. This analysis 
implies that  the kinetic freeze-out, where elastic particle-particle 
collisions cease, cannot be occurring at a condition very different 
from the chemical freeze-out.  However, one pion
HBT analysis at $p_\bot<0.5$ GeV suggests
kinetic pion freeze-out at about $T_k\simeq120$ MeV \cite{Fer99}.
A possible explanation of why here considered $p_\bot>0.7$ GeV
particles are not subject to a greater spectral deformation
after chemical freeze-out, is  that they escape before the bulk 
of softer hadronic particles is formed.

The six statistical parameters describing the  particle abundances
are shown in the top section of table~\ref{fitqpbs}, for both 
Pb- and S-induced reactions \cite{LRa99}.  
The  errors  shown are one standard  deviation errors arising
from the propagation of the experimental measurement
error, but apply only when the  theoretical model describes 
the data well, as is the case here, see  the header of each column
--- note that for the S-induced reactions (see  \cite{LRa99})  the number 
of redundancies is large since same data comprising
different kinematic cuts is included in the analysis. 
We note the interesting result that within error the freeze-out 
temperature $T_{\rm f}$ seen in table~\ref{fitqpbs},
is the same for both the S- and Pb-induced reactions. 
The collective velocity rises from
$v_c^{\rm S}=0.5c$ to $v_c^{\rm Pb}\simeq c/\sqrt{3}=0.58$.
We then show the light quark fugacity $\lambda_{q}$,
and  note $\mu_B^{\rm Pb}=203\pm5 >\mu_B^{\rm S}=178\pm5$\,MeV.
As in  S-induced reactions where $\lambda_{s}=1$,
now  in Pb-induced reactions, a value $\lambda_{s}^{\rm Pb}\simeq 1.1$
characteristic for a source of freely movable  strange quarks with
balancing strangeness, {\it i.e.,} $\tilde\lambda_{s}=1$ is obtained, 
see Eq.\,(\ref{lamQ}). 

$\gamma_q>1$ seen in table~\ref{fitqpbs} implies  that there is
phase space over-abundance of light quarks, to which,
{\it e.g.,} gluon fragmentation at QGP breakup {\it prior} to hadron 
formation contributes.  $\gamma_q$ assumes in our data 
analysis a value near to  where pions 
could begin to  condense\cite{Heinz99},  
$\gamma_q=\gamma_q^c\equiv e^{m_\pi/2T_f}$\,. 
We found studying the ratio $h^-/B$
separately from other experimental results
that the value of $\gamma_q\simeq\gamma_q^c$ is fixed
consistently and independently both, by the negative hadron ($h^-$),
and the strange hadron yields. The unphysical  range 
$\gamma_q>\gamma_q^c$ can  arise, since up to this
point we use only a first quantum (Bose/Fermi) 
correction. However, when  Bose distribution for 
pions is implemented, which requires  the 
constraint $\gamma_q\le\gamma_q^c$, 
we obtain  practically the same results, as shown
in second column of table~\ref{fitqpbs}. Here
we allowed only 4 free parameters, {\it i.e.} 
we set $\gamma_q=\gamma_q^c$\,, and  the strangeness
conservation constraint fixes $\lambda_s$\,. 
We then show in table~\ref{fitqpbs} the ratio $\gamma_s/\gamma_q$,
which corresponds (approximately)  to the parameter $\gamma_s$ when
$\gamma_q=1$ had been  assumed.  We note  that $\gamma_s^{\rm Pb}>1$.
This strangeness over-saturation effect could arise from the effect 
of gluon fragmentation combined with early chemical equilibration
in QGP, $\gamma_s(t<t_f)\simeq 1$. The ensuing rapid expansion
preserves this high strangeness yield, and thus we find the result
$\gamma_s>1$\,, as is shown in figure 33 in \cite{acta96}.

We show in the bottom section of table~\ref{fitqpbs} the
energy and entropy content per baryon,
and specific anti-strangeness content,
along with specific strangeness asymmetry 
of the hadronic particles emitted.
The  energy  per baryon seen in the emitted hadrons is nearly
equal to the available specific energy
of the collision  (8.6 GeV for Pb--Pb, 8.8--9 GeV for S--Au/W/Pb).
This implies that the fraction of energy deposited in the central
fireball  must be nearly the same as the fraction of baryon number.
The small reduction of the specific entropy in Pb--Pb compared to
the lighter S--Au/W/Pb system maybe driven by the greater baryon
stopping in the larger system, also seen in the smaller energy per
baryon content. Both collision systems freeze out at energy per unit
of entropy $E/S=0.185$ GeV. 
There is a loose relation of this universality in the 
chemical freeze-out condition with the suggestion made
recently that particle freeze-out occurs at a fixed energy per baryon for
all physical systems \cite{CR98}, since the entropy content is related to
particle multiplicity. The overall high specific entropy content we find
agrees well with the entropy content
evaluation made  earlier \cite{Let93} for the S--W case.

Inspecting figure 38 in \cite{acta96} we see
that the specific yield of strangeness we expect
from the kinetic theory in QGP is at the level of 0.75 
per baryon, in agreement with the results of
present analysis shown in table \ref{fitqpbs}. 
This high strangeness yield leads
to the enhancement of multi-strange (anti)baryons,
which are viewed as important hadronic signals of
QGP phenomena \cite{Raf80}, and a series of recent
experimental analysis has carefully demonstrated  comparing
p--A with A--A results that
there is quite significant enhancement \cite{WA97,WA85}, 
as has also been noted before by the experiment NA35 \cite{Alb94}.

The strangeness imbalance seen in the asymmetrical S--Au/W/Pb system
(bottom of table  \ref{fitqpbs}) could be a real effect arising from 
hadron phase space properties. However, this result
also reminds us that though the statistical errors are very small, 
there could be a considerable
systematic error due to presence of other contributing
particle production mechanisms. Indeed, we do not offer here 
a consistent understanding of the  $\Omega,\,\overline\Omega$ yields
which are higher than we can describe. 
We have explored additional  microscopic mechanisms. 
Since the missing $\Omega,\,\overline\Omega$ yields are
proportional (13\%) to the  $\Xi,\overline\Xi$ yield, we have  tested  
the  hypothesis of string fragmentation contribution in  the 
{\it final state}, which  introduces  just the needed `shadow' of the 
$\Xi,\overline\Xi$ in the $\Omega,\overline\Omega$ 
abundances. While this works for $\Omega,\,\overline\Omega$, 
we find that this mechanism is not compatible with the other 
particle abundances.  We have also explored the 
possibility that unknown $\Omega^*,\,\overline{\Omega^*}$
resonances contribute to the $\Omega,\,\overline\Omega$ yield, but
this hypothesis is ruled out since the missing yield is clearly 
baryon--antibaryon asymmetric. Thus though we reached here a very 
good understanding of other hadronic particle yields and spectra, the 
rarely produced but greatly enhanced  $\Omega,\,\overline\Omega$ 
must arise in  a more complex hadronization pattern. 

We have presented a comprehensive analysis of hadron
abundances and $m_\bot$-spectra  observed in Pb--Pb 158$A$ GeV interactions
within the statistical Fermi model with chemical non-equilibrium 
 of strange and non-strange hadronic particles.  
The key results we obtained  are: $\tilde \lambda_s=1$ for
S and Pb collisions\,; $\gamma_s^{\rm Pb}>1, \ \gamma_q>1$\,;
 $S/B\simeq 40$\,; $ s/B\simeq 0.75$\,;
all in a remarkable agreement with the properties of a deconfined
QGP source hadronizing  without chemical re-equilibration, and 
expanding not faster than the sound velocity of quark matter. 
The universality of the physical properties
at chemical freeze-out for S- and Pb-induced
reactions points to a common nature of the
primordial source  of hadronic particles in both systems. The difference
in spectra between the two systems arises in our analysis from the difference
in the collective surface explosion velocity, which for larger system 
is higher, having  more time to develop.
Among other interesting results which also verify the consistency
of our approach  we recall: good  balancing of
strangeness $\langle \bar s-s\rangle=0$ in the Pb--Pb case;
increase of the baryochemical potential as the
collision system grows;
energy per baryon near to the value expected if energy and baryon number
deposition in the fireball are similar. We note that 
given the magnitude of $\gamma_q$ and low chemical freeze-out
temperature, most (75\%) of all  final state pions are 
directly produced, and not resonance decay products.
 Our results differ significantly
from an earlier analysis regarding the temperature at which hadron formation
occurs. Reduction to $T_f=140$--$145$\,MeV  becomes possible 
since we allow for the chemical  non-equilibrium and collective flow, 
and only with these improvements in analysis 
our description acquires convincing statistical significance,
which e.g. a  hadronic gas scenario does not offer \cite{BHS99}. 
Because we consider flow 
effects, we can address the central rapidity data of the WA97 
experiment at the required level of precision,  
showing the consistency in these results with the 
NA49 data discussed earlier \cite{Bec98}. 

In our opinion, the only consistent interpretation of 
the experimental results analyzed here is
that hadronic particles seen at 158$A$ GeV 
nuclear collisions at CERN-SPS
are  formed directly in hadronization of an exploding  
deconfined  phase of hadronic matter,
and that these particles do not undergo a 
chemical re-equilibration after they have been
produced.  

We  thank U. Heinz, E. Quercigh, J. Sollfrank and R.L. Thews 
for valuable comments. 
Supported in part by a grant from the U.S. Department of
Energy,  DE-FG03-95ER40937\,. LPTHE, Univ.\,Paris 6 et 7 is:
Unit\'e mixte de Recherche du CNRS, UMR7589.
\vskip -0.6cm


\begin{table}[tb]
\caption{\label{resultpb2}
WA97 (top) and NA49 (bottom)  Pb--Pb 158$A$ GeV particle ratios
and our theoretical results, see text for explanation.}
\small\begin{center}
\begin{tabular}{|lcl|ll|ll|}
\tableline
 Ratios & $\!\!\!\!$Ref. &  Exp.Data                                  & All & All$|_v$&  No-$\Omega$ & No-$\Omega|_v$  \\
\tableline
${\Xi}/{\Lambda}$ &\cite{Kra98} &0.099 $\pm$ 0.008                     &  0.107  & 0.110   &  0.095  & 0.102   \\
${\overline{\Xi}}/{\bar\Lambda}$ &\cite{Kra98} &0.203 $\pm$ 0.024      &  0.216  & 0.195   &  0.206  & 0.210   \\
${\bar\Lambda}/{\Lambda}$  &\cite{Kra98} &0.124 $\pm$ 0.013            &  0.121  & 0.128   &  0.120  & 0.123   \\
${\overline{\Xi}}/{\Xi}$  &\cite{Kra98} &0.255 $\pm$ 0.025             &  0.246  & 0.225   &  0.260  & 0.252   \\
${\Omega}/{\Xi}$      &\cite{Kra98} &0.192 $\pm$ 0.024                 &  0.192  & 0.190   &0.078$^*$&0.077$^*$\\
${\overline{\Omega}}/{\overline{\Xi}}$  &\cite{Hol97} &0.27 $\pm$ 0.06 &  0.40   & 0.40    &0.17$^*$ &0.18$^*$ \\
${\overline{\Omega}}/{\Omega}$  &\cite{Kra98} &0.38 $\pm$ 0.10         &  0.51   & 0.47    &0.57$^*$ &0.60$^*$ \\
$(\Omega+\overline{\Omega})\over(\Xi+\bar{\Xi})$&\cite{Hol97}&0.20 $\pm$ 0.03
                                                                       &  0.23   & 0.23    &0.10$^*$ &0.10$^*$ \\
\tableline
$(\Xi+\bar{\Xi})\over(\Lambda+\bar{\Lambda})$&\cite{Ody97}  &0.13 $\pm$ 0.03
                                                                      &  0.109  & 0.111   &  0.107  & 0.114   \\
${K^0_{\rm s}}/\phi$   &\cite{Puh98}  & 11.9 $\pm$ 1.5\ \       &  16.1   & 15.1    &  9.89   & 12.9    \\
${K^+}/{K^-}$         &\cite{Bor97}         &  1.80$\pm$ 0.10         &  1.62   & 1.56    &  1.76   & 1.87    \\
$p/{\bar p}$     &\cite{Ody98}              &18.1 $\pm$4.\ \ \ \      &  16.7   & 15.3    &  17.3   & 17.4    \\
${\bar\Lambda}/{\bar p}$     &\cite{Roh97}  & 3. $\pm$ 1.             &  0.65   & 1.29    &  2.68   & 2.02    \\
${K^0_{\rm s}}$/B       &\cite{Jon96}       & 0.183 $\pm$ 0.027       &  0.242  & 0.281   &  0.194  & 0.201   \\
${h^-}$/B                 &\cite{Jon96}     & 1.83 $\pm $ 0.2\ \      &  1.27   & 1.55    &  1.80   & 1.83    \\
\tableline
 &   &                                      $\chi^2_{\rm T}$          &  19     & 18      &  2.1    & 1.8\\
 &   &                                      $ N;p;r$                  &15;5;4   & 16;6;4  & 11;5;2  & 12;6;2\\
\tableline
\end{tabular}
\end{center}
\vskip -0.8cm
\end{table}

\begin{table}[bt]
\caption{\label{Tetrange2}
Experimental and theoretical $m_\bot$ spectra inverse slopes $T_{\rm th}$.
Left Pb--Pb results 
from experiment NA49 \protect\cite{Mar99}
for  kaons and from experiment WA97 \protect\cite{WA97} for baryons;
right  S--W  results from  WA85  \protect\cite{WA85slopes}.}
\begin{center}
\begin{tabular}{|l|cc|cc|}
\tableline
               & $T_{\bot}^{\rm Pb}$\,[MeV]&$T_{\rm th}^{\rm Pb}$\,[MeV]&$T_{\bot}^{\rm S}$\ [MeV]&
                                                                   $T_{\rm th}^{\rm S}$\ [MeV]\\
\tableline
$T^{{\rm K}^0}$             & 223 $\pm$  13&  241& 219 $\pm$  \phantom{1}5 &  215\\
$T^\Lambda$                 & 291 $\pm$  18&  280& 233 $\pm$  \phantom{1}3 & 236\\
$T^{\overline\Lambda}$      & 280 $\pm$  20&  280& 232 $\pm$  \phantom{1}7 & 236\\
$T^\Xi$                     & 289 $\pm$  12&  298& 244 $\pm$  12& 246\\
$T^{\overline\Xi}$          & 269 $\pm$  22&  298& 238 $\pm$  16& 246\\
\tableline
\end{tabular}
\end{center}
\vskip -0.8cm
\end{table}

\begin{table}[b!]
\caption{\label{fitqpbs}
Top section: statistical parameters, and their $\chi^2_{\rm T}$, 
which best describe the experimental results for
Pb--Pb data, and in last column for S--Au/W/Pb data presented in Ref. \protect\cite{LRa99}\,.
Bottom section: specific energy, entropy, anti-strangeness, net strangeness
 of  the full hadron phase space characterized by these
statistical parameters. In the middle column we fix $\lambda_s$ by requirement of 
strangeness conservation and choose $\gamma_q=\gamma_q^c$, the pion condensation point.}
\vspace{-0.2cm}\begin{center}
\begin{tabular}{|l|cc|c|}
\tableline
                       &  Pb--No-$\Omega|_v$& Pb--No-$\Omega|_v^*$ & S--No-$\Omega|_v$ \\
$\chi^2_{\rm T};\ N;p;r$&1.8;\ 12;\,6;\,2   & 4.2;\ 12;\,4;\,2 &  6.2;\ 16;\,6;\,6 \\
\tableline
$T_{f}$ [MeV]          &    144 $\pm$ 2     &  145 $\pm$ 2    &  144 $\pm$ 2 \\
$v_c$                  &   0.58 $\pm$ 0.04  & 0.54 $\pm$ 0.025 &   0.49 $\pm$ 0.02\\
$\lambda_{q}$          &   1.60 $\pm$ 0.02  & 1.605 $\pm$ 0.025 &   1.51 $\pm$ 0.02 \\
$\lambda_{s}$          &   1.10 $\pm$ 0.02  & 1.11$^*$         & 1.00 $\pm$ 0.02   \\
$\gamma_{q}$           &   1.7 $\pm$ 0.5    & $\gamma_q^c=e^{m_\pi/2T_f}$    &   1.41 $\pm$ 0.08 \\
$\gamma_{s}/\gamma_{q}$&   0.86 $\pm$ 0.05  & 0.78 $\pm$ 0.05 &  0.69 $\pm$ 0.03  \\
\tableline
$E_{f}/B$              &   7.0 $\pm$ 0.5  & 7.4 $\pm$ 0.5   &  8.2 $\pm$ 0.5    \\
$S_{f}/B$              &    38 $\pm$ 3    &    40 $\pm$ 3 &   44 $\pm$ 3     \\
${s}_{f}/B$            &  0.78 $\pm$ 0.04 & 0.70 $\pm$ 0.05 &   0.73 $\pm$ 0.05 \\
$({\bar s}_f-s_f)/B$   &  0.01 $\pm$ 0.01 & 0$^*$ &    0.17 $\pm$ 0.02\\
\tableline
\end{tabular}
\end{center}
\vskip -0.8cm
\end{table}

\end{narrowtext}

\end{document}